\documentclass[10pt,twocolumn]{article} 
\usepackage{simpleConference}
\usepackage{times}
\usepackage{graphicx}
\usepackage{amssymb}
\usepackage{url,hyperref}
\usepackage{subcaption}
\usepackage{graphicx}
\usepackage{subcaption}
\usepackage{mathtools}
\usepackage{xcolor}
\usepackage{siunitx}
\usepackage{booktabs}
\usepackage{siunitx,etoolbox}
\usepackage{amsmath}
\usepackage{textcomp}

\DeclareMathOperator*{\argmax}{argmax}

\begin{document}

\title{Lossless Compression of Mosaic Images with Convolutional Neural Network
	Prediction}

\author{Seyed Mehdi Ayyoubzadeh\\
	McMaster University\\
	Hamilton, ON\\
	{\tt\small ayyoubzs@mcmaster.ca}
	\and
	Xiaolin Wu\\
	McMaster University\\
	Hamilton, ON\\
	{\tt\small xwu@mcmaster.ca}
}

\maketitle
\thispagestyle{empty}

\begin{abstract}
	
	We present a CNN-based predictive lossless compression scheme for raw color mosaic
	images of digital cameras. This specialized application problem was previously understudied
	but it is now becoming increasingly important, because modern CNN methods for image
	restoration tasks (e.g., superresolution, low lighting enhancement, deblurring), must operate
	on original raw mosaic images to obtain the best possible results. The key innovation of this
	paper is a high-order nonlinear CNN predictor of spatial-spectral mosaic patterns. The deep
	learning prediction can model highly complex sample dependencies in spatial-spectral mosaic
	images more accurately and hence remove statistical redundancies more thoroughly than
	existing image predictors. Experiments show that the proposed CNN predictor achieves
	unprecedented lossless compression performance on camera raw images.
	
\end{abstract}

\section{Introduction}

\begin{figure}
	\centering
	\includegraphics[width=0.70\linewidth]{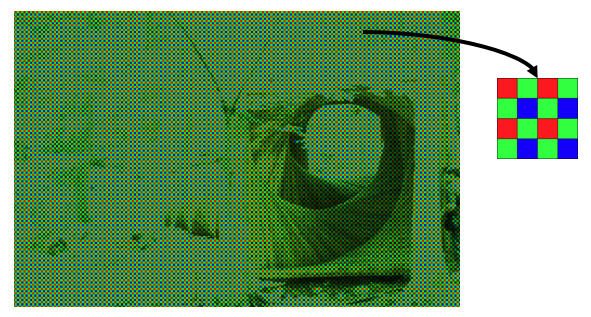}
	\caption{Bayer pattern example \cite{DangNguyen2015}}
	\label{fig:bayerPattern}
\end{figure}

The vast majority of digital cameras sample image signals into a 2D spatial and spectral mosaic through a color filter array (CFA) as shown in Figure \ref{fig:bayerPattern}. The camera output image is the result of processing raw mosaic data through a digital image processing pipeline (DIP) that is integrated into the camera chip. DIP performs a sequence of operations on mosaic data:
denoising, demosaicing, white balance, gamma correction, quantization and compression to generate an RGB image, usually in JPEG format. But these DIP operations incur information losses and most of them are irreversible. While DIP offers an inexpensive real-time mainstream solution to color image generation, it by no means offers the best possible image quality. Recently, the CNN-based methods for image restoration tasks, such as superresolution,
deblur, low lighting reconstruction, etc., beat all previous methods of explicit modeling and achieve unprecedented restoration performances \cite{NIPS2012_4824, 1512.03385, 1703.06870, 1708.02002}. But in order to produce the best
results, these convolutional neural networks need to take the original mosaic image data as input; in other words, any information loss caused by DIP has negative impact on the final inference result. For this reason, professional SLR cameras and many of high end smartphones offer a lossless mode, in which users can access to raw mosaic sensor readings. However, if not compressed, storing raw mosaic images is not practical due to their sheer size. This gives rise to the subject of this paper, the lossless compression of CFA mosaic images.
The above stated problem has been hardly studied in any depth. Technically, losslessly compressing mosaic images is much more difficult than lossless compression of conventional images, because the spatial-spectral interleaving of pixels generates multiway intricate sample correlations. The traditional and most effective method for lossless image compression is predictive coding, also known as differential pulse coding modulation (DPCM)
\cite{Tomar2015}. In DPCM, pixel samples are sequentially predicted and the prediction residuals are entropy coded. The compression performance of the DPCM system primarily depends on the precision of the predictor. Previous studies on predictors for lossless image compression were mostly devoted to the linear type. This is motivated not by the physical nature of the problem but rather by the computational amenity of designing linear predictors. Classical linear predictors for image coding can be found in \cite{XiaolinWu1, XinLi,DBLP:journals/tip/AkimovKF07, Memon1995}
Although linear prediction is effective to decorrelate stationary Gaussian random processes, it is ill suited for natural ismages that have nonstationary statistics, and even more so for color mosaic images.
In this work, we develop a deep convolutional neural network (DCNN) predictor for lossless compression of color mosaic images. This mosaic image prediction network, called MIPnet, breaks away from the linearity limitations of traditional image predictors. Indeed, deep learning can, in theory, model complex, non-linear causal relationships of spatially interleaved spectral samples, provided that a large amount of paired input and output data is available. Unlike in many image restoration tasks for which the ground truth images are unavailable or
very hard and expensive to have, the good news for the construction of MIPnet is that the required training image pairs are readily available and practically unlimited.
In previous predictive image compression methods, the prediction is carried out sequentially on a pixel by pixel basis. The pixel-by-pixel traversal is operationally an awkward way, if not impossible, of predicting color mosaic images because different pixels are samples of different spectral bands. Even more problematically, scanning an image one pixel at a time obscures the original two-dimensional pixel structures. In the design of MIPnet, a two-dimensional block of mosaic pixels, not a single pixel, are predicted as a coding unit of the
lossless compression system. Our new design has two advantages: 1. the spatial-spectral mosaic configuration is treated as a whole; 2. the block prediction strategy introduces an implicit regularization mechanism to prevent overfitting of the network predictor.

\subsection{Related Work}

In \cite{Koh} Koh and Mitra published a method for compressing Bayer mosaic images. They transform the image with methods so-called structure separation and structure conversion respectively. After transformations, they compress images using JPEG. Their compression method is not lossless since they use JPEG compression.  Lee and Ortega in \cite{SangYongLee} proposed a simple approach for CFA images compression. First, they convert the RBG color space to YCbCr, then they rotate the Y channel and fill the micro blocks. Eventually, they compress each channel by JPEG algorithm. In \cite{BattiatoAFV} Battiato et al.proposed a method based on vector quantization followed by an entropy encoder for CFA images compression. \\ In \cite{NingZhang2006} Zhang et al. proposed a method for lossless compression of Color mosaic images. They used Mallat packet wavelet for direct compression of mosaic images without de-interleaving. Xie et al. proposed a method in which they first apply a low pass filter and then perform $2:1$ down-sampling on the green channel. Then, they use RGB to YCbCr color conversion to use JPEG lossy compression \cite{XiangXie}. Chang and Chun in \cite{KingHongChung2008} proposed a prediction-based lossless compression of CFA images. They separate each CFA image into two sub images, the green subimage and nongreen subimage. Then, the green subimage is encoded by using context matching based prediction and the nongreen channel with using the color difference between the green and non green subimages. Sonehara et al. in \cite{Sonehara1989} proposed a method in which they used Multi Layer Perceptrons (MLP) for image compression.
As previously mentioned, DCNNs can be used as the predictors in the DPCM encoding scheme. In fact, any generative model can be used as the predictor. Generative models divided into two general categories: i) Auto-Regressive models ii) Auto Encoders and Generative Adversarial Networks (GANs).
Auto-regressive models have several advantages over non auto regressive models:
a) auto-regressive models can provide an approach to explicitly compute likelihood, this makes them more suitable for compression. b) Training of these networks are more stable than trainig GANs. c) auto-regressive models work both for discrete and continous data.  

\begin{figure}
	\centering
	\includegraphics[width=\linewidth]{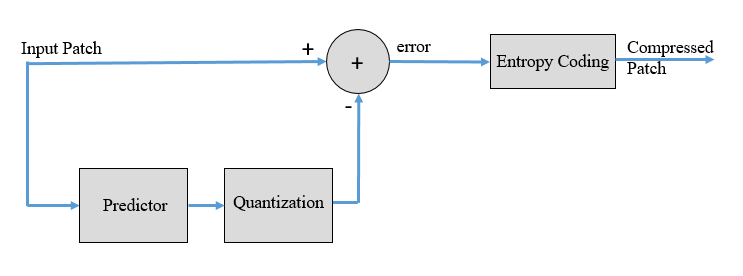}
	\caption{Block diagram of lossless compression}
	\label{fig:dpcmBlock}
\end{figure}

Probably, the most popular examples of auto-regressive models are PixelRNN \cite{1601.06759} and PixelCNN \cite{1606.05328}. The main disadvantage of the PixelRNN is that it is quite slow in training and in inference time. Although PixelCNN is faster than PixelRNN, it predicts R, G and B channels sequentially. Consequently, it is not suitable for Bayer mosaic images.  
\begin{figure*}[t]
	\centering
	\begin{subfigure}[b]{0.4\linewidth}
		\includegraphics[width=\linewidth]{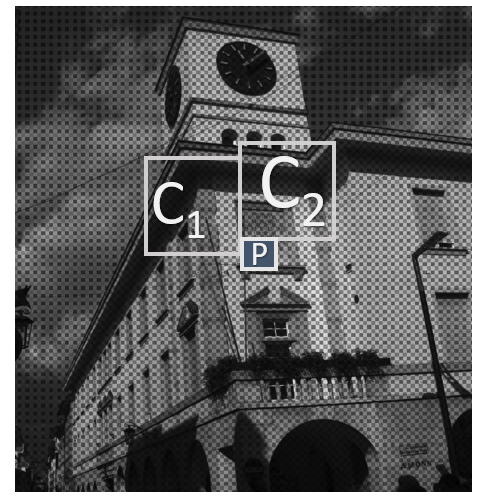}
	\end{subfigure}
	\begin{subfigure}[b]{0.45\linewidth}
		\includegraphics[width=\linewidth]{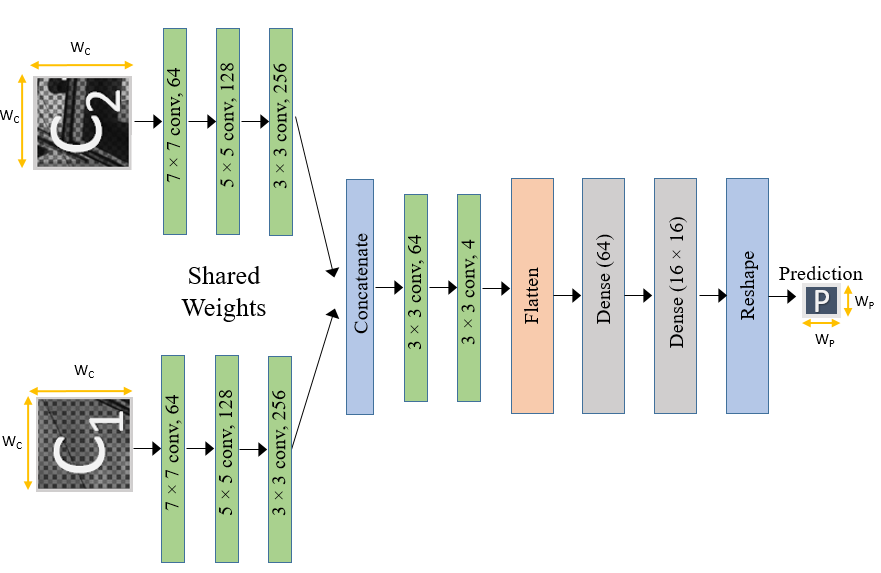}
	\end{subfigure}
	\caption{(a) Contexts of a patch in a Bayer pattern image (b) Feature extraction from contexts for patch prediction}
	\label{fig:prediction_pipeline}
\end{figure*}

Wu et. al used CNN as a non linear predictor to extrapolate images \cite{1808.07757}. Similarly Gong and Yang in \cite{Gong2017VideoFI} proposed a method to use CNN as a non linear predictor for video frame interpolation and extrapolation. Shi et al. propose a method in \cite{1506.04214} for precipitation nowcasting. They changed the operations in LSTM cells to convolutional operations to use them on images. In \cite{Chen2018} Chen and He used DCNNs for stock prediction. In \cite{Toderici2017} Toderici et al. published a method for image compression using Recurrent Neural Networks (RNNs). They used two RNNs as encoder and decoder, a binarizer and a neural network for entropy coding. \\
Ahanonu in \cite{LosslesssCompressionMaster} published a method to use DCNNs as  predictors of Discrete Wavelet Transform (DWT) subbands. They used DCNNs as the predictors of different wavelet transformed images. Besides the fact that this method is computationally expensive, it can not be applied directly to the mosaic images. In \cite{1611.01704} Ballé et al. proposed a technique for lossy compression using a two-stage DCNN. They optimized the DCNN with minimizing a weighted sum of rate and distortion under the assumption of fixed uniform scaler quantizer in the code space. Zhou et al. used variational autoencoders for image compression. They improved the compression performance further by utilizing pyramidal feature fusion structure at the encoder \cite{Zhou2018VariationalAF}. In \cite{HernandezCabronero2018}, Cabronero et. al change JPEG-2000 compression pipeline for mosaic images.
In \cite{1705.05823} Rippel and Bourdev used GANs for lossy compression with a outstanding compression ratio. Similarly, Agustsson et al. in \cite{1804.02958} used GANs for image compression for extremely low bit rates. These methods are in the category of lossy compression and can not be applied directly to raw images. In \cite{1811.12817} Mentzer et al. used a non auto regressive model for lossless compression of RGB images.

\section{Proposed Method}

\subsection{Lossless Compression}
Assume we have a set of symbols $X=\{x_1,...,x_n\}$ and there is a stream of the symbols $x$, which are independent and identically distributed (i.i.d) according to the probability mass function $p_{sym}$. The goal of the lossless encoder is to compress this stream such that the expected bits per symbol is minimized $L=\sum_{i=1}^{n}p_{sym}(x_i)l(x_i)$, where $l(x_i)$ is the length of the symbol $x_i$ and the stream can be recovered in the decoder. Note that, information theory ensures us $L$ is always greater than the Shannon entropy of $p_{sym}$ ($H(p_{sym})=\sum_{i=1}^{n}-p_{sym}(x_i)log(p_{sym(x_i)})$) \cite{Shannon1948}.

\subsection{Adaptive Arithmetic Coding}
Arithmetic coding can provide us a nearly optimal data compression if it is used in conjuction with suitable probabilistic model for $p_{sym}$ \cite{Howard1992}. In encoding images, the pixels have strong spatial correltion and so they are not i.i.d samples. In this case, we can use the chain rule in probability to factorize the probability distribution of the image $p_{img}$ to $p_{img}(x)=\prod_{k} p(x_k|x_{k-1},\cdots,x_1)$. Now, to encode $x$ we can use $x_k$ symbol and encode that with the probability distribution $p(x_k|x_{k-1},\cdots,x_1)$. Generally this approach is known as Adaptive Arithmetic Coding (AAC). Note, that the factorization must be causal since receiver uses $p_{img}$ and it varies in every step \cite{1811.12817}. 

\subsection{Overview of the Encoder}
The encoder pipeline is shown in Figure \ref{fig:dpcmBlock}. First, the input image is split into $W_P \times W_P$ non-overlapping patches. The predictor (DCNN) tries to predict each patch using its contexts as shown in Figure \ref{fig:prediction_pipeline} (a). To have a causal encoder, the chosen contexts must be in the left or top of the patch (previous pixels of the patch). The predicted patch is quantized and bounded afterwards. Then, it is subtracted from the ground truth patch to obtain the error map (residual). Eventually, the erro map is encoded via arithmetic coding. \\

\subsection{Calculating Compression Ratio} \label{compression_ratio}
One of the most important steps to have a fair comparison with other methods is to compute compression ratio realistically. Typically, source coders implement context modeling to derive an arithmetic source coder \cite{XiaolinWu}. 
In the case of $1D$ example, if $(X_1, X_2, \cdots, X_N)$ denote the sequential data, $J$ previous samples of $X_i$ can be employed as the context $C_i$ of $X_i$ ($C_i=(X_{i-1}, \cdots, X_{i-J})$). $P(X_i|C_i)$ can be estimated by its histogram. \\ However, for large number of symbols, the estimation may have a considerable error \cite{XiaolinWu}. To resolve this issue, pioneering methods for image compression quantize the context and then estimate $P(X_i|Q(C_i))$, where $Q(C_i)$ is a context quantizer.  \\  Assume the required number of bits for uncompressed image is $B$. The compression ratio is defined as follows:

\begin{equation} 
CR=\frac{B}{H(E|Q(C))}
\end{equation}
Where $E$ is the error map. Note the error images are mosaic when we compress mosaic images. 

\begin{figure}
	\centering
	\includegraphics[width=\linewidth]{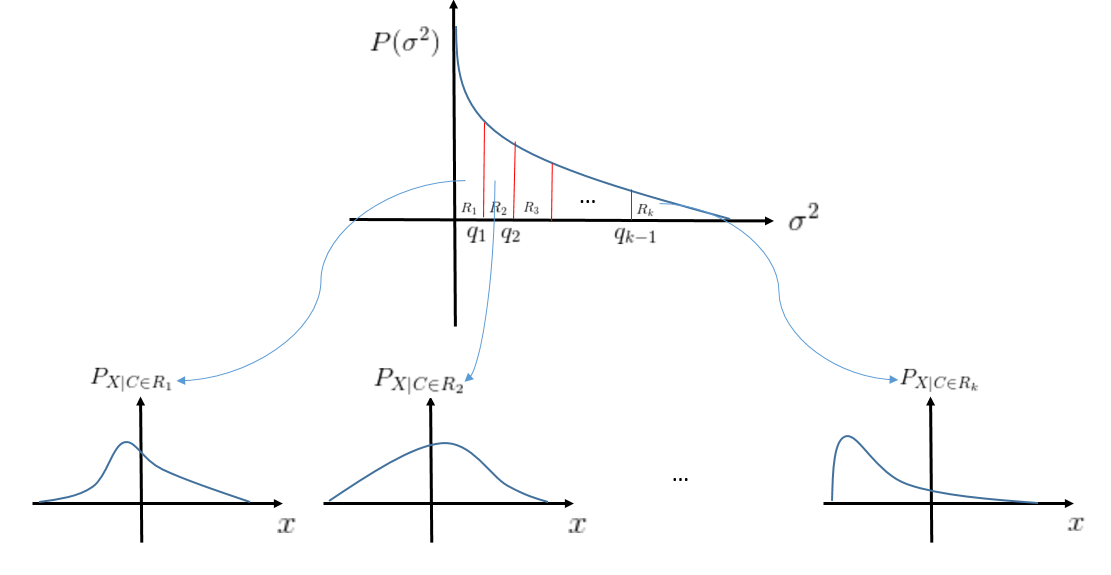}
	\caption{The procedure of designing optimal quantizer}
	\label{fig:quantizer}
\end{figure}

\subsection{Designing the Optimal Context Quantizer}
\begin{figure}
	\centering
	\includegraphics[width=\linewidth]{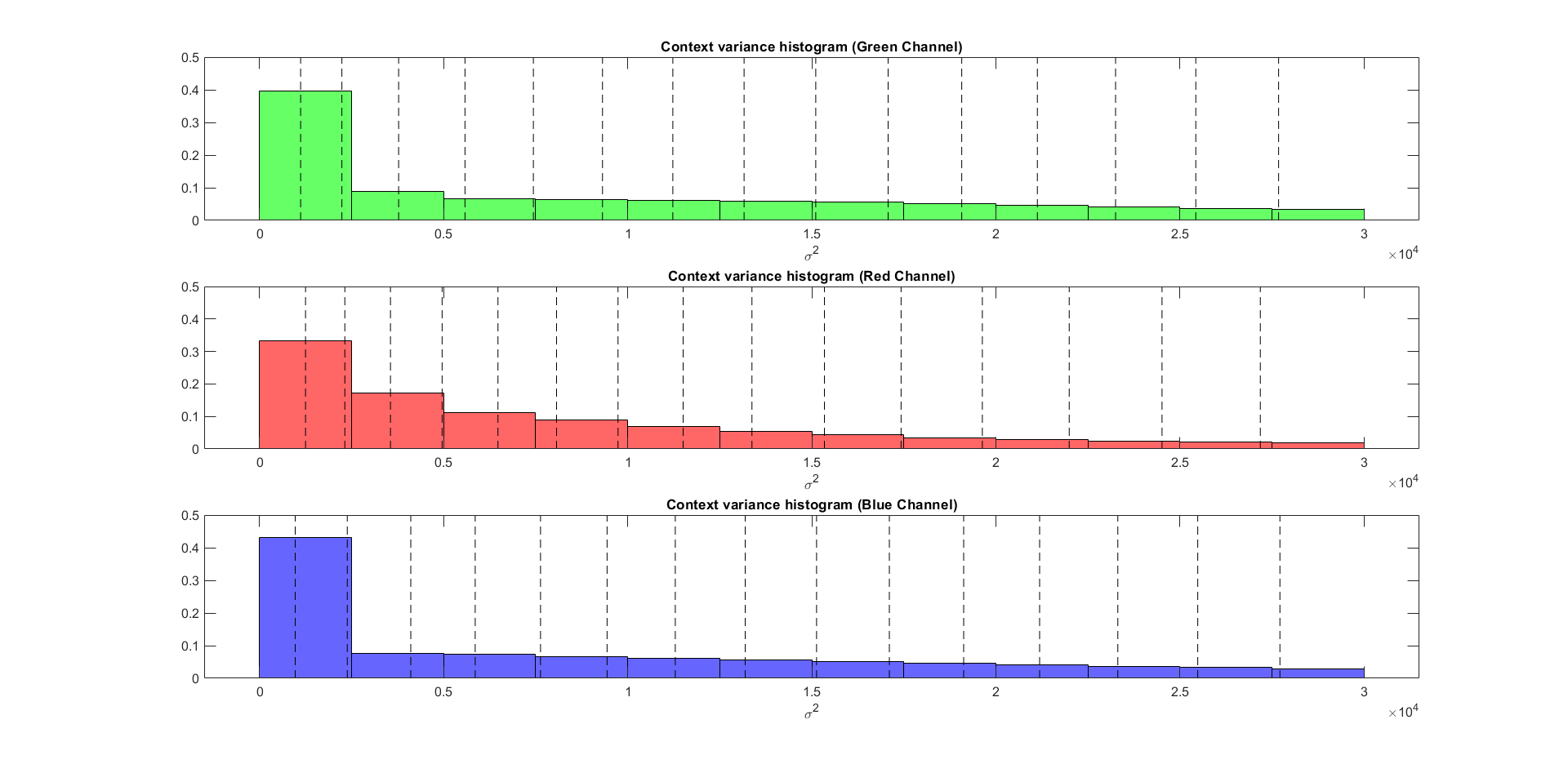}
	\caption{Context variance hisogram for different channels (the quantizer paritions are shown with dashed lines for $K=16$)}
	\label{fig:context_var}
\end{figure}

\begin{figure}
	\centering
	\includegraphics[width=\linewidth]{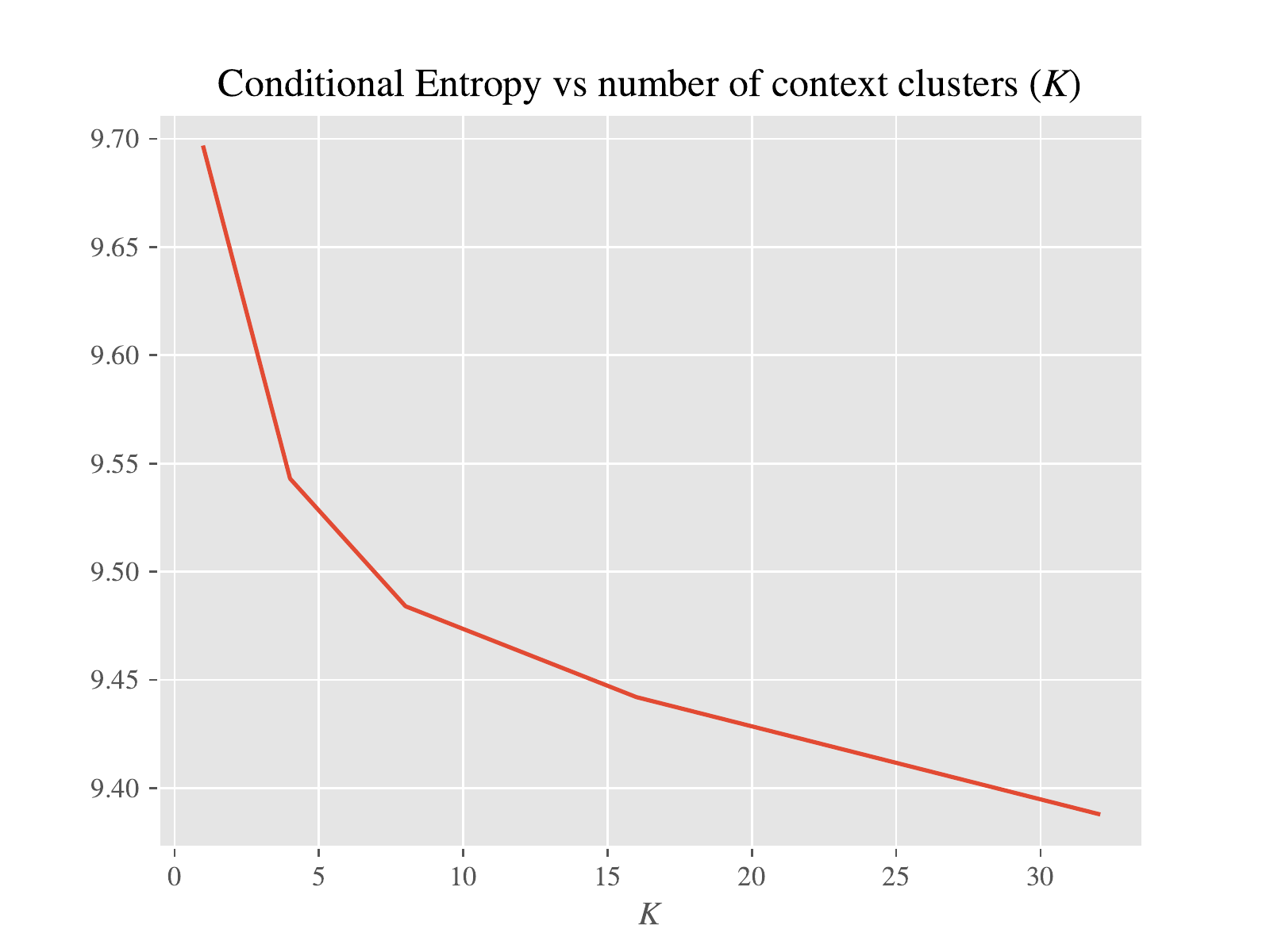}
	\caption{Conditional entropy vs $K$}
	\label{fig:entropy}
\end{figure}

We need a feature for context modeling to design the arithmatic coder. Variances (energies) of the pixels are suitable features since they can divide the pixels into a handful of categories. Next, a quantizer is needed to quantize the variances. For each pixel, we use a $3 \times 3$ patch to obtain the patch variance. Data Preprocessing Inequality (DPI) ensures $H(X|Q(C)) \geq H(X|C)$. Therefore, the quantizer should be designed to minimize the $H(X|Q(C)) - H(X|C)$. This is the Kullback-Leibler divergence between the probability distribution of $P(X|C)$ and the distribution after the quantization ($P(X|Q(C))$). We have depicted how qunatizer splits the signal into different regions in Figure \ref{fig:quantizer}. The goal of the optimal quantizer is to partition the variances to $K$ different regions such that the resulting bit rate ($\sum_{i=1}^{K} -P(C_i)log(P(X|C_i))$) is minimized.
This problem can be solved by Dynamic programming \cite{XiaolinWu}. In this paper, to quantize the variances we employed Lloyd quantization algorithm which minimizes the reconstruction distortion \cite{Lloyd1982}. $10$ images have been used to design the quantizer. We have shown the variances histograms for different channels in Figure \ref{fig:context_var} for $K=16$. The conditional entropy versus number of regions ($K$) is shown in Figure \ref{fig:entropy}. $K=16$ is the suitable number of regions for the quantizer.

Concerning the error images are mosaic, we need to separate the error image into $4$ different channels ($E_R, E_B, E_{G_1}, E_{G_2}$). Under the assumption that the channel probabilities are equal, the condtional entropy can be computed as follows:

\begin{align}
H(E|Q(C)) &= \frac{1}{4}H(E_R|Q(C_R)) + \frac{1}{4}H(E_B|Q(C_B)) \\ &+ \frac{1}{4}H(E_{G_1}|Q(C_{G_1})) + \frac{1}{4}H(E_{G_2}|Q(C_{G_2})) \nonumber
\end{align}
Given $K$ categories, we can compute conditional entropies by using the corresponding histograms. \\ 

\subsection{Zero-frequency issue}
One problem that can lead to have an over optimistic estimation of the compression ratio is zero frequency issue. When the entropy is calculated via histograms, a large number of bins may have zero frequency occurrences. In such case, entropy estimation provides an optimistic estimation of the entropy, however in the test time, the zero-frequency bins symbols can occur and decrease the bit rate. We weight the non zero frequency bins by a factor $\alpha$ ($\alpha >> 1$) then add $1$ occurrence to all the bins to address this issue.  $\alpha$ can emphasize on the bins in the histogram that already exist. Assume $N_{sym}$ and $N_{tot}$ are the total number of possible bins in the histogram (e.g., for $8$ bit images $N_{sym}=256$) and the sum of the occurrences respectively. The probability of symbol $i$ can be calculated as:

\begin{equation}
P_i=\begin{cases}
\frac{\alpha n_i + 1}{\alpha N_{tot}+N_{sym}}, & \text{i is a non-zero-frequency bin}.\\
\frac{1}{\alpha N_{tot}+N_{sym}}, & \text{i is a zero-frequency bin}
\end{cases}
\end{equation}

\begin{figure}
	\begin{center}
	\includegraphics[width=\linewidth]{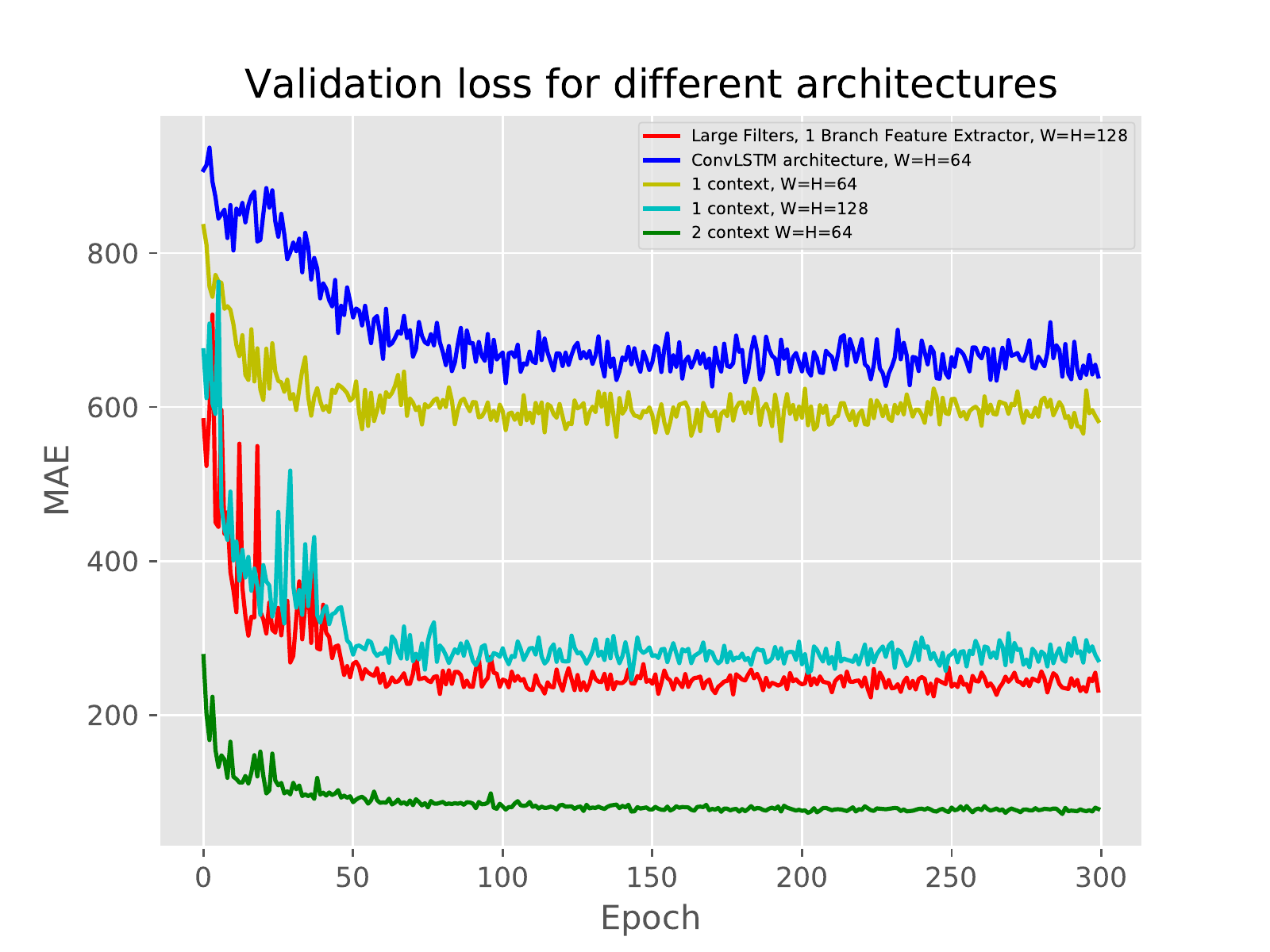}
	\caption{Validation loss for different architectures}
	\label{fig:val_loss_diff}		
	\end{center}
\end{figure}
Where, $n_i$ is the frequency of $i$th bin. The resulted entropy of this histogram is:
\begin{align}
H = - \sum_{i=1}^{N_{sym}} P_ilog_2(P_i)
\end{align} 

This method is a remedy for the zero-frequency problem in the entropy estimation and gives a better estimation when histograms are used. 

\subsection{Predictor}
\subsubsection{DCNN Architecture}
DCNN is the non linear predictor that predicts each patch by its surrounding contexts. The prediction pipeline is shown Figure \ref{fig:prediction_pipeline}.
Two surrounding contexts of each patch are used for the prediction to give the DCNN (predictor) more information about spatial structure of the patch.  
The extracted contexts are not spatially aligned, therefore stacking them channelwise is not the best choice for feeding them to the DCNN. Note that feeding the contexts within a large square and zero padding in the prediction patch position can intorduce artifacts for the DCNN, thus this is not a proper way to feed the contexts to the DCNN. Instead, we pass the contexts through two different branches of convolution layers that have shared weights to extract features from both of them while maintaining spatial structures as illustrated in the Figure \ref{fig:prediction_pipeline} (b). We have tried various architectures, including ConvLSTM \cite{1506.04214} layers, stacking contexts as different channels and using different number of contexts. The best architecture based on our experiment is shown \ref{fig:prediction_pipeline} (b). We use relatively large kernel size for the convolutional filters of the DCNN. In DCNNs with many convolutional layers, using large filter size is not necessary, since in the last convolutional layers the receptive field of the kernels are large enough. However, in DCNNs with few number of convolutional layers for having sufficient receptive field we should use large filter sizes. Our porposed DCNN architecture is shown in Figure \ref{fig:prediction_pipeline} (b). We call our proposed DCNN MIPnet. MIPnet has only $1,017,988$ parameters. In comparison with the DCNNs that perform image enhancement and similar tasks on the mosaic images, MIPNet has considerably lower computational complexity and number of layers. Therefore, MIPNet can be implemented practically on the integrated circuits.      

\subsection{Loss Function of MIPNet}
By determining the architecture of the MIPNet. Now, we should define a loss function for the DCNN. The Laplacian distribution is widely used to model the residual statistics. We can use Maximum Likelihood (ML) estimation to use this probabilistic information about the residuals. Due to the periodic structure of Bayer mosaic images, it is reasonable to predict a $2 \times 2$ patch in each step. Assume the vectorized input is an $n$ dimensional vector. The vectorized output has $4$ dimensional. Therefore, the regression problem is find a mapping from input space ${\rm I\!R}^n$ to the corresponding output space ${\rm I\!R}^4$. We assume that outputs are independent from each other. Let $\boldsymbol{\hat{y}}$ and $\boldsymbol{y}$ denote the prediction of MIPNet and the ground truth for each component, respectively. Under Laplacian distribution assumption for the residuals ($\boldsymbol{e}$), for each component of $\boldsymbol{e}$ ($e_i$) we have:

\begin{align} 
& e_i(\boldsymbol{x}) =  y_i - \hat{y_i}(\boldsymbol{x}) \nonumber \\ 
& e_i(\boldsymbol{x})  \sim Laplace(0, b) \\
& y_i = \hat{y}(\boldsymbol{x}) + e(\boldsymbol{x})  \nonumber\\
& y_i  \sim Laplace(\hat{y_i}(\boldsymbol{x}), b) \nonumber 
\end{align}

Where $b > 0$ is the diversity of Laplacian distribution and $\boldsymbol{x}$ is the input of the DCNN ($\boldsymbol{x} \in {\rm I\!R}^n$). For simplicity, we replace $b$ with $1$ since it does not change the behavior of the loss function. The training dataset contains $N$ pairs of datapoints denoted by $(\boldsymbol{x_1}, \boldsymbol{y_1}), \cdots ,(\boldsymbol{x_N}, \boldsymbol{y_1})$, in which $\boldsymbol{x_i}$s and $\boldsymbol{y_i}$s are 1-D vectors. Now, we are able to estimate $\boldsymbol{\hat{y}}$ using ML estimation. It is the mean of the conditional distribution of $\boldsymbol{\hat{y}}$ given $\boldsymbol{x}$. Therefore, we can write:
\begin{align} 
\boldsymbol{\hat{y}}(\boldsymbol{x}) &= \argmax_{\boldsymbol{\hat{y}}(\boldsymbol{x})} \sum_{i=1}^{N}
log(P(\boldsymbol{y}^{(i)}))
\label{eqn:ml}
\end{align}
Substituting the conditional distribution of $\boldsymbol{y}$ given $\boldsymbol{x}$   into \ref{eqn:ml} yields to the following loss function:
\begin{align} 
L_M &= -\sum_{i=1}^{N}
log(P(\boldsymbol{y}^{(i)}(\boldsymbol{x})) = \frac{1}{N} \|\boldsymbol{y}-\boldsymbol{\hat{y}}\|_1
\end{align}
\subsubsection{Context Regularization}
To encourage the DCNN to capture spatial structure and patterns in the output, instead of predicting $2 \times 2$ blocks for the outputs, we predict $2S \times 2S$ patches, where $S$ is an positive integer. This is in fact a certain type of Multi Task Learning (MTL). MTL has several advantages including, attention focusing, representation bias of the DCNN and regularization of the DCNN \cite{1706.05098}. In fact, the extra information about the neighborhood of the output patch can help the DCNN for the prediction, however it can not pass it to the DCNN as the input. Because, it violates the casuality of the DCNN in the inference time. This trick let MIPNet to use the extra information as the auxiliary output while keeping the causality of the DCNN in the inference time.  
In our experiments we have used the scale factor of $S=8$. Note that, in the inference time, we only utilize the first $2 \times 2$ block as predictor output. In addition, since different color channels have different statistics, it is necessary to decompose context regularization loss for each channel. Therefore, context regularization loss can be written as follows:

\begin{align*} 
L_{CR} = \frac{1}{N} \sum_{i=1}^{N} (\lambda_g   \|\boldsymbol{y_{a,g}}-\boldsymbol{\hat{y}_{a,g}}\|_1 + \\
\lambda_r \|\boldsymbol{y_{a,r}}-\boldsymbol{\hat{y}_{a,r}}\|_1 +
\lambda_b  \|\boldsymbol{y_{a,b}}-\boldsymbol{\hat{y}_{a,b}}\|_1)
\end{align*}

\begin{figure}
	\centering
	\includegraphics[width=0.75\linewidth]{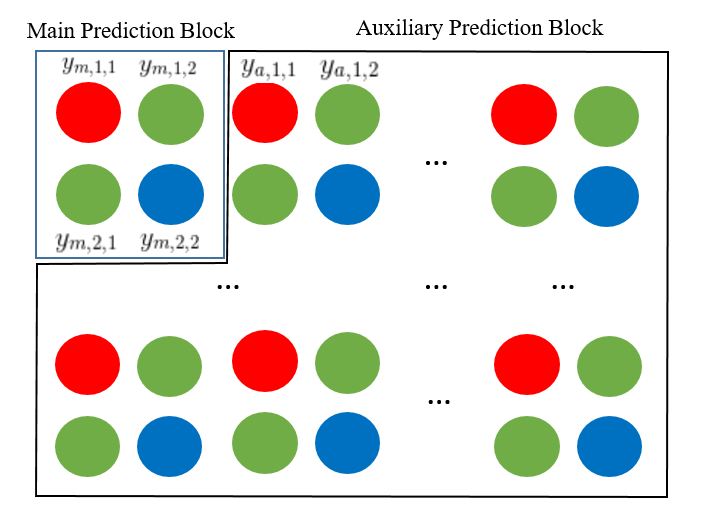}
	\caption{Main and Auxiliary Pixels}
	\label{fig:aux_prediction}
\end{figure}

Where $\boldsymbol{y_{a,ch}}$ and $\boldsymbol{\hat{y}_{a,ch}}$ are the auxiliary output pixels of the ground truth and the DCNN for the channel ($ch \in \{r,g,b\}$)  respectively. $N$ is the total number of auxiliary pixels and  $\lambda_r$, $\lambda_g$ and $\lambda_b$ denoting the context regularization factors for different color channels. We have demonstrated the main prediction pixels and auxiliary pixels in the Figure \ref{fig:aux_prediction}.
The total loss function of MIPNet is the sum of main loss of the DCNN ($L_M$) and context regularization loss ($L_{CR}$). MIPNet tries to minimize this loss function concerning its parameters.

\section{Experiments}
We evaluate all of the experiments on RAISE-6K dataset \cite{DangNguyen2015}. This dataset consists of $4276$ 14-bit NEF (Nikon Electronic Format) images. 

\subsection{Different Architectures}
Different architectures are tested to choose the right architecture for MIPNet. The validation accuracy for different architectures is shown in Figure \ref{fig:val_loss_diff}. As one can see extracting two contexts for each prediction patch with the architecture shown in Figure \ref{fig:prediction_pipeline} (b) has the best performance in comparison with other architectures. In fact, choosing the right architecture is critical for this task and have significant impact on the performance of the DCNN. It can be seen that architecture with ConvLSTM has lower performance in comparison with others. The reason is the vanishing gradient problem for the LSTM layer.

\subsection{Effect of the Context Regularization Loss}
\begin{figure}
	\centering
	\includegraphics[width=\linewidth]{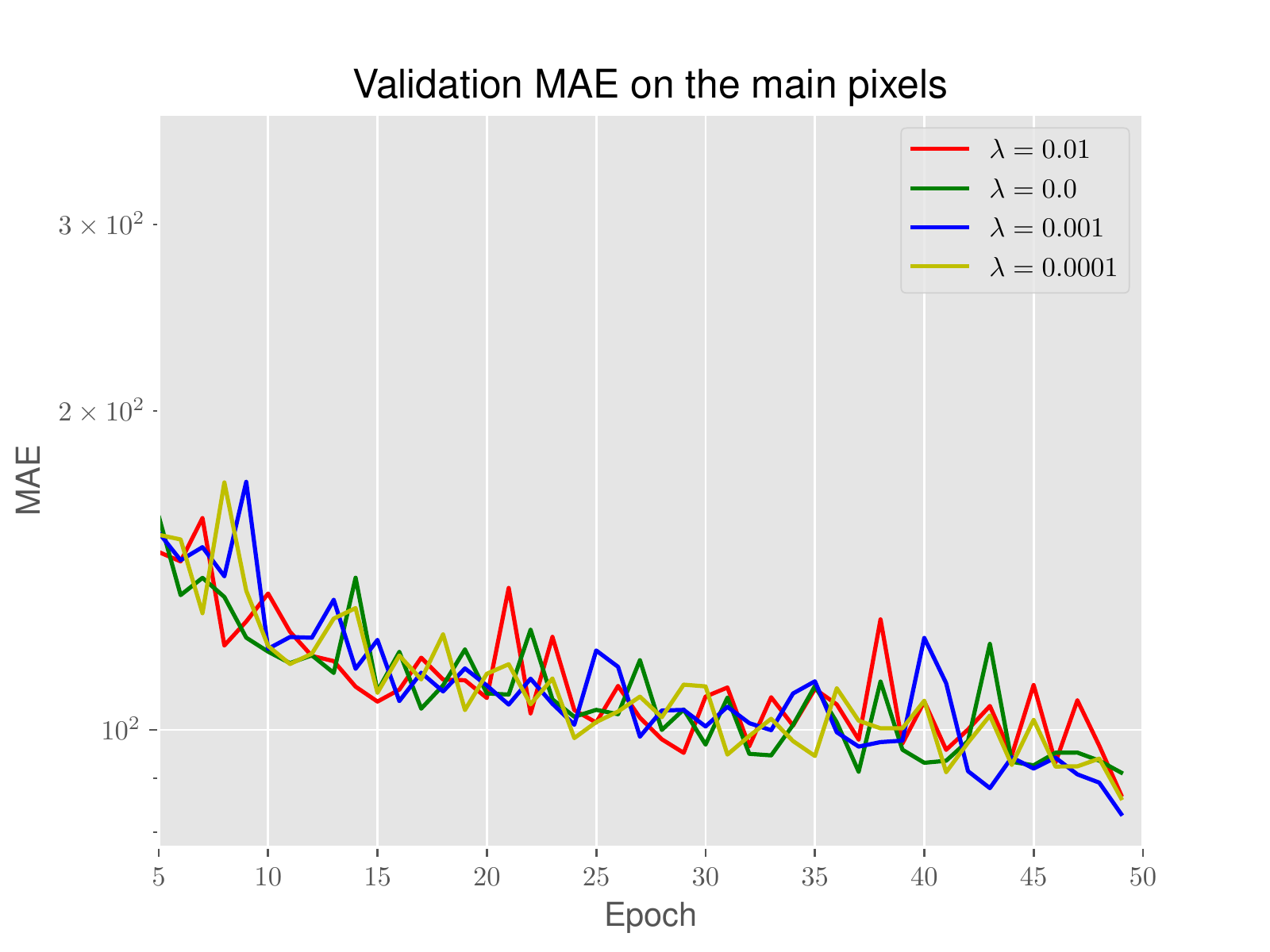}
	\caption{MAE on the prediction pixels for different $\lambda$s}
	\label{fig:diff_lambdas}
\end{figure}

To show the effectiveness of the context regularization loss, we train the MIPNet with different context regularization loss factors. The validation loss is shown in Figure \ref{fig:diff_lambdas}.

As one can see when the DCNN does not have context regularization loss ($\lambda =0$), the performance is lower in comparison when the DCNN uses the extra information provided by auxiliary pixels.

\begin{figure}
	\centering
	\includegraphics[width=\linewidth]{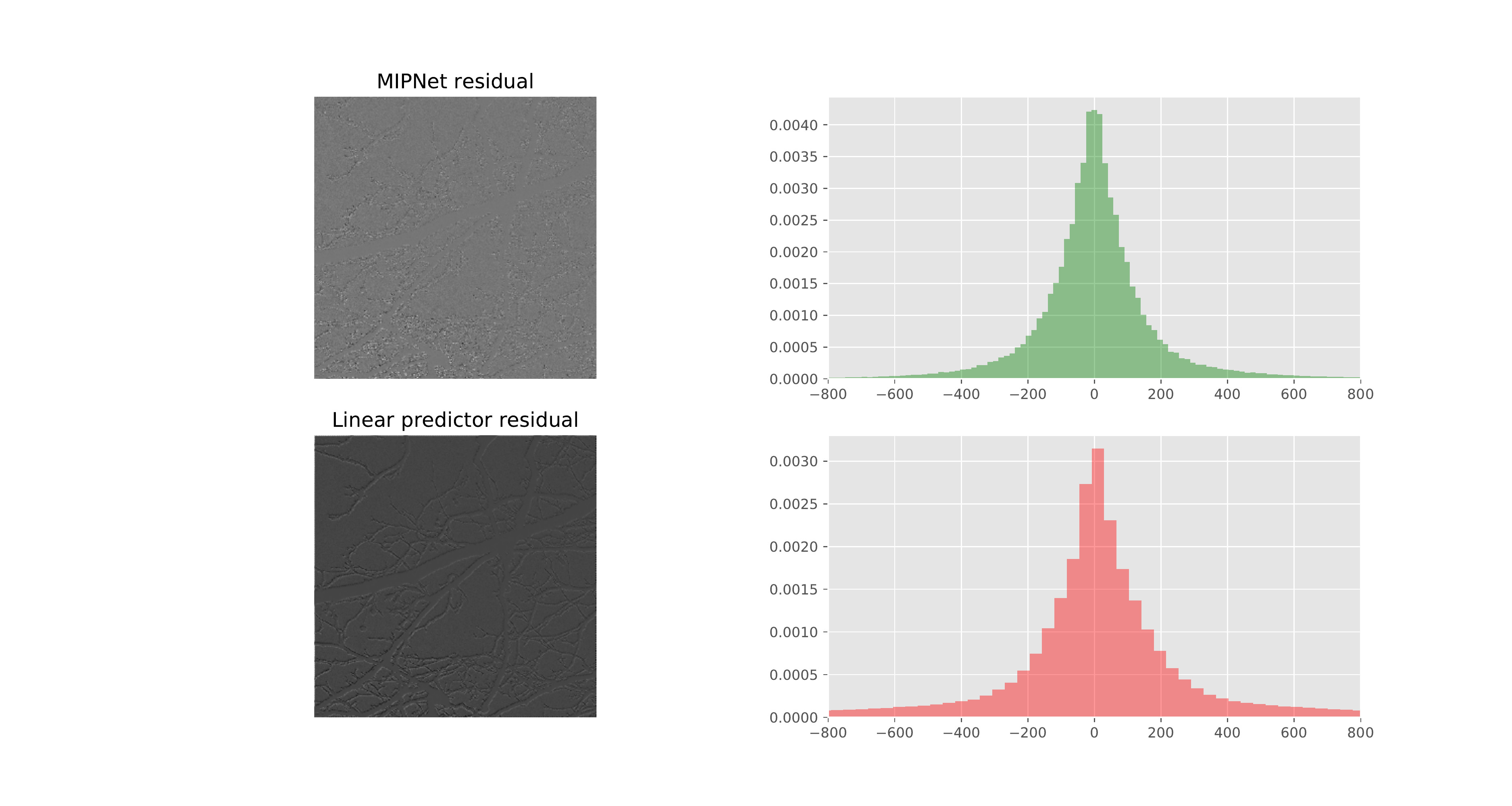}
	\caption{The residual maps for MIPNet and a linear predictor with correspoing histograms}
	\label{fig:test_image_res}
\end{figure}

\begin{figure}
	\centering
	\includegraphics[width=0.6\linewidth]{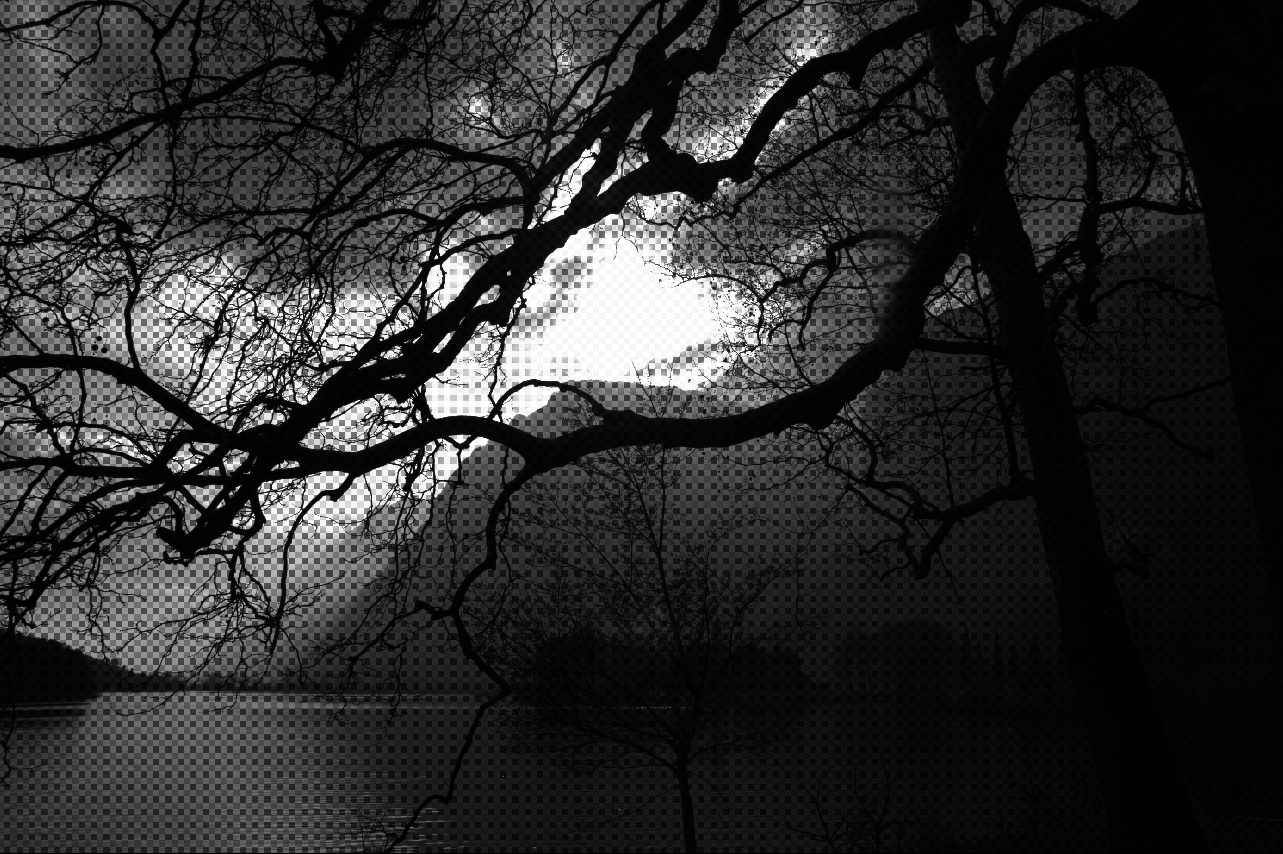}
	\caption{Test image}
	\label{fig:test_image}
\end{figure}

\begin{figure}
	\centering
	\includegraphics[width=\linewidth]{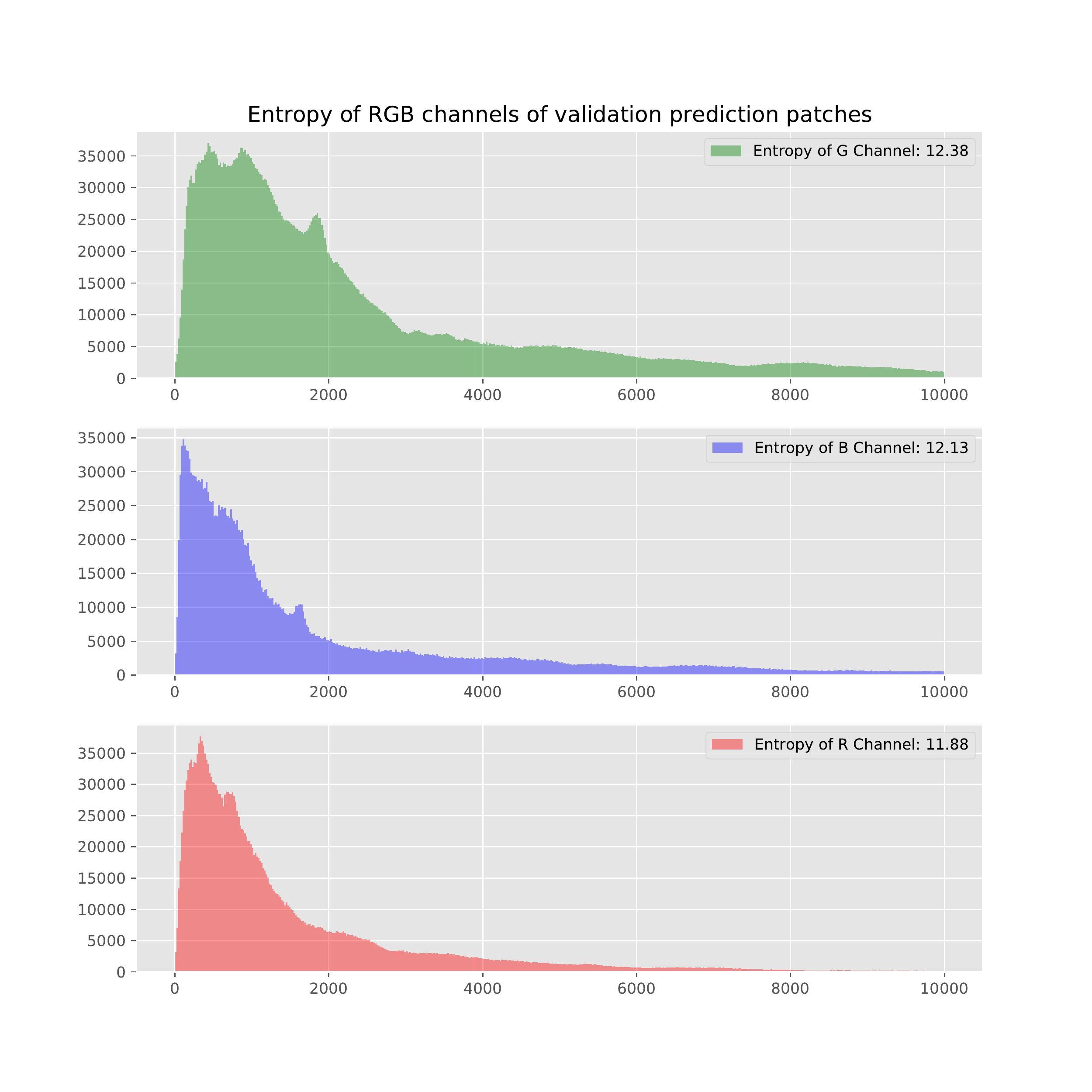}
	\caption{Histogram of the validation patches}
	\label{fig:hists_val}
\end{figure}

\begin{figure}
	\centering
	\includegraphics[width=\linewidth]{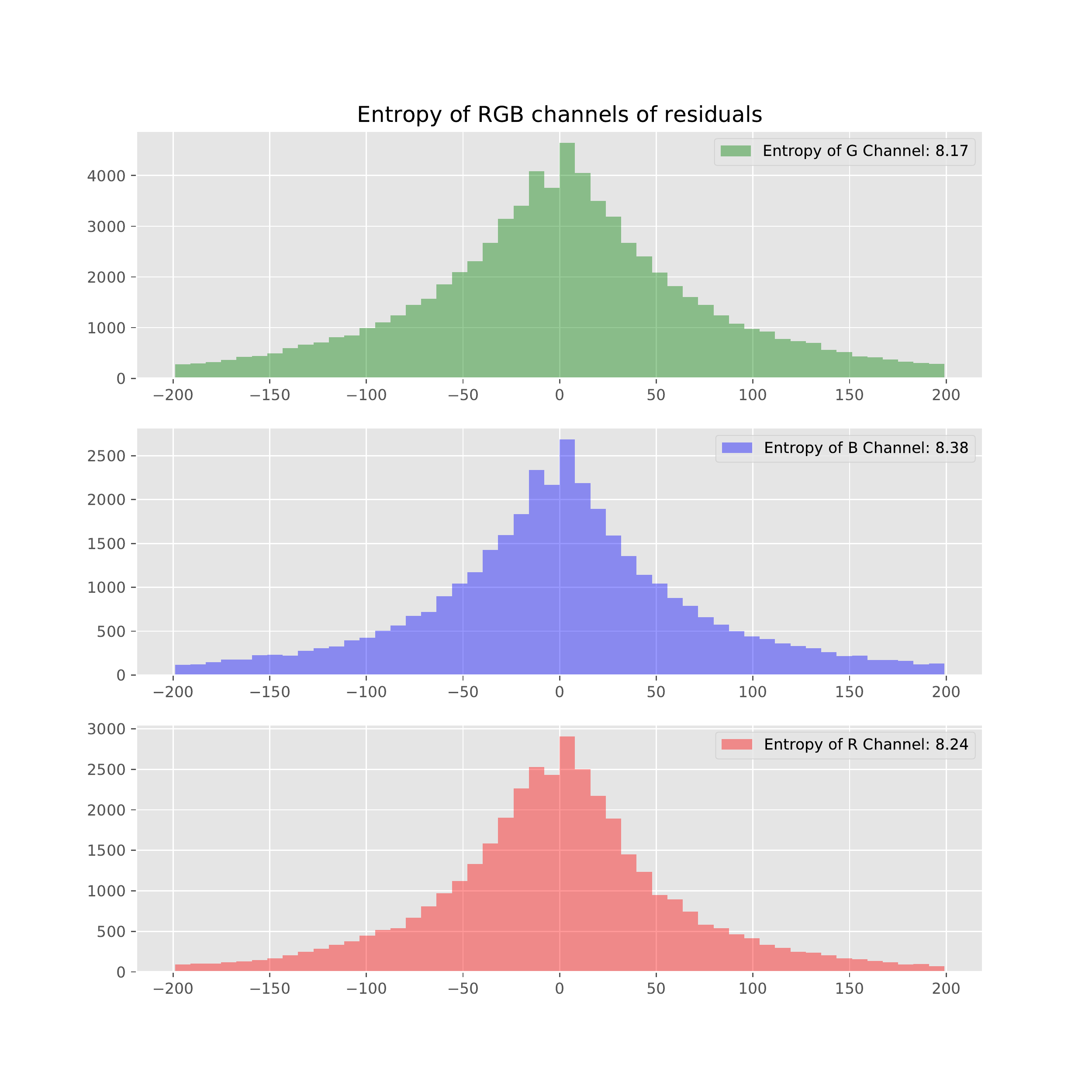}
	\caption{Histogram of the error patches}
	\label{fig:hists_err}
\end{figure}

\subsection{Training Setup}
We have used Leaky ReLU \cite{1505.00853} activation functions for all units. The size of contexts ($W_c \times W_c$) is $64 \times 64$. We have trained the model for $300$ epochs and scaled the learning rate by $0.4$ after each $50$ epochs. Initial value of the learning rate is $10^{-4}$. We set $\lambda_b$, $\lambda_g$ and $\lambda_r$ to $10^{-3}$. 

\begin{figure}
	\centering
	\includegraphics[width=\linewidth]{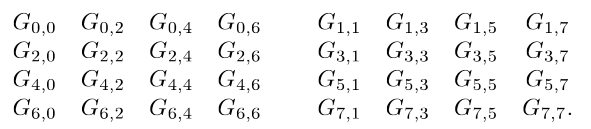}
	\caption{Illustration of separation of g channels \cite{NingZhang2006}}
	\label{fig:separation}
\end{figure}

\begin{figure}
	\centering
	\includegraphics[width=0.5\linewidth]{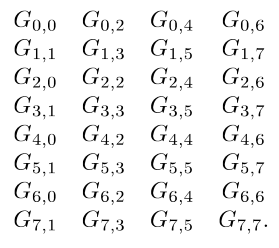}
	\caption{Illustration of merging of g channels \cite{NingZhang2006}}
	\label{fig:merge}
\end{figure}

\subsection{Error Histograms}
It is beneficial to visualize the error histogram for the validation data and compare that with the validation data histogram. As shown in Figure \ref{fig:hists_err}, the error histogram for validation patches can be well approximated by Laplace distribution, and in comparison with \ref{fig:hists_val}, the entropy is significantly lower. Note that in the case of 14-bit images, the range of pixels intensities is from $0$ to $(2^{14}-1)=16383$. We have evaluated our method for both 14 bit and 8 bit images. As shown in Figure \ref{fig:hists_val} and \ref{fig:hists_err}, the compression of the DCNN for the green channel is higher in comparison with R and B channels. This is due to the fact that, in the contexts there are more green pixels and therefore more information for green channels. To calculate this compression ratio, we need to calculate the conditional entropy of residuals, as explained in section \ref{compression_ratio}. A test image and its corresponding residual image for MIPNet and a linear predictor are shown in Figures \ref{fig:test_image} and \ref{fig:test_image_res}. The residual image shows that the DCNN can predict the blocks well except for very high frequency parts of the image, and in comparison with the linear predictor, the residual histogram for MIPNet is sharper and has a smaller variance.

\subsection{Performance on 14 bit and 8 bit images}
In order to show the effectiveness of MIPNet, we compare our method with JPEG-2000, PNG and FLIF which are the most popular lossless compression methods. Since we have two different G channels we perform merge and separation methods for JPEG-2000 on G channels. merge and separation procedures for G channels are shown in Figure \ref{fig:separation} and \ref{fig:merge}. The results are shown in Table \ref{tab:table1}. As one can see, our method has the lowest lossless bit rate in comparison with other methods.

\definecolor{ao(english)}{rgb}{0.0, 0.5, 0.0}

\begin{table*}
	\begin{center}
		\caption{Lossless bit rates of 14 bit mosaic test images (\textcolor{ao(english)}{\textit{green}} percentages show the improvement regarding to the PNG compression)}
		\label{tab:table1}
		\begin{tabular}{l|c|c|c|c|c}
			\toprule 
			\textbf{Image} & \textbf{PNG} & \textbf{JPEG2K-Separation} & \textbf{JPEG2K-Merge} & \textbf{FLIF} & \textbf{Ours} \\
			\midrule 
			Tree & 11.61 & 9.78 & 9.68 &  9.04 & 9.30  \\
			Statue & 10.89 & 8.10 & 8.08 & 7.90 &  7.71  \\
			Bike & 11.91 & 9.65 & 9.62 &  9.41 & 8.78   \\
			Monument & 11.38 & 9.98 & 9.83 & 8.79 & 8.75  \\
			Grass & 11.53 & 9.09 & 9.07 & 8.67 & 7.55  \\
			Artwork & 10.96 & 8.91 & 8.87 & 7.86 & 7.46  \\
			\small Mountain Snow & 11.43 & 10.06 & 9.92 & 8.66 & 8.27  \\
			Street & 11.29 & 8.10 & 8.08 & 8.16 & 8.47 \\
			Cow & 11.55 & 8.83 & 8.82 & 8.78 & 8.66  \\
			\bottomrule 
			Average & 11.39 & 9.16 (\textcolor{ao(english)}{19.5\%}) & 9.10 (\textcolor{ao(english)}{20.1\%})  & 8.62 (\textcolor{ao(english)}{24.3\%})  & $\mathbf{8.29}$ (\textcolor{ao(english)}{$\mathbf{27.2}$\%}) \\ 			
			\bottomrule 
		\end{tabular}
	\end{center}
\end{table*}

To compare our method with the other state of the art engineered codec WebP \cite{43447}, we have to evaluate our method on $8$ bit images since WebP does not support $14$ bit images compression. The results are shown in Table \ref{tab:table2}.
\begin{table*}
	\begin{center}
		\caption{Lossless bit rates of 8 bit mosaic test images (\textcolor{ao(english)}{\textit{green}} percentages show the improvement regarding to the PNG compression)}
		\label{tab:table2}
		\begin{tabular}{l|c|c|c|c|c|c}
			\toprule 
			\textbf{Image} & \textbf{PNG} & \textbf{JPEG2K-Separation} & \textbf{JPEG2K-Merge} &  \textbf{WebP} & \textbf{FLIF} & \textbf{Ours} \\
			\midrule 
			Tree & 4.3 & 3.95 & 3.87 & 3.90 & 3.39 & 3.67 \\
			Statue & 2.72 & 2.28 & 2.27 &  2.19 & 1.98 & 2.00 \\
			Bike & 3.91 & 3.57 & 3.54 &  3.53 & 3.37 & 3.10 \\
			Monument & 3.54 & 3.15 & 3.12 &  3.08 & 2.92 & 3.19 \\
			Mountain & 3.42 & 2.99 & 2.96 &  2.94 & 2.71 & 2.53 \\
			Artwork & 3.65 & 3.07 & 3.05 &  3.00 & 2.82 & 2.59 \\
			\small Mountain Snow & 4.18 & 3.84 & 3.82 &  3.76 & 3.74 & 3.46 \\
			Street & 3.18 & 2.73 & 2.72 &  2.70 & 2.52 & 2.49 \\
			Cow & 3.43 & 2.94 & 2.93 &  2.83 & 2.74 & 2.70 \\
			\bottomrule 
			Average & 3.59 & 3.16 (\textcolor{ao(english)}{11.9\%}) & 3.14 (\textcolor{ao(english)}{12.5\%}) &  3.10 (\textcolor{ao(english)}{13.6\%}) & 2.91 (\textcolor{ao(english)}{18.9\%}) & $\mathbf{2.85}$ (\textcolor{ao(english)}{$\mathbf{20.6}$\%})  \\ 
			
			\bottomrule 
		\end{tabular}
	\end{center}
\end{table*}

As we can see our method outperforms other methods for 8 bit mosaic images.

\section{Conclusion}
In this paper, we proposed the very first method for lossless compression of mosaic images via DCNNs. We have shown that our method outperforms other traditional methods for lossless compression, such as JPEG-2K, FLIF, WebP and PNG of mosaic images. This method is ideally suited in applications that require preserving high fidelity images after compression. Furthermore, our proposed DCNN is a light weight model computational wise. Consequently, this method can be practically employed by integrated circuits in the cameras before digital image processing modules to compress raw images concurrently.

\clearpage

\bibliographystyle{abbrv}
\bibliography{compression_for_review}
\end{document}